\documentclass[%
reprint,
superscriptaddress,
 amsmath,amssymb,
 aps,
 prl,
floatfix,
]{revtex4-2}

\usepackage{graphicx}
\usepackage{dcolumn}
\usepackage{bm}
\usepackage{hyperref}
\usepackage[mathlines]{lineno}

\begin{document}

\preprint{APS/123-QED}

\title{Electron Heating by Parallel Electric Fields in Magnetotail Reconnection}

\author{Louis Richard}
\email{louis.richard@irfu.se}
\affiliation{
Swedish Institute of Space Physics, Uppsala 751 21, Sweden
}%

\author{Yuri V. Khotyaintsev}
\affiliation{
Swedish Institute of Space Physics, Uppsala 751 21, Sweden
}%

\author{Cecilia Norgren}
\affiliation{
Swedish Institute of Space Physics, Uppsala 751 21, Sweden
}%

\author{Konrad Steinvall}
\affiliation{Department of Physics, Chalmers University of Technology, G\"oteborg 412 96, Sweden
}%

\author{Daniel B. Graham}
\affiliation{
Swedish Institute of Space Physics, Uppsala 751 21, Sweden
}%

\author{Jan Egedal}
\affiliation{
Department of Physics, University of Wisconsin-Madison, Madison, Wisconsin 53706, USA
}%

\author{Andris Vaivads}
\affiliation{Division of Space and Plasma Physics, KTH Royal Institute of Technology, Stockholm 100 44, Sweden}
\affiliation{Ventspils University of Applied Sciences, Ventspils 3601, Latvia}

\author{Rumi Nakamura}
\affiliation{Space Research Institute, Austrian Academy of Sciences, Graz 8042, Austria}

\date{\today}

\begin{abstract}
We investigate electron heating by magnetic-field-aligned electric fields ($E_\parallel$) during anti-parallel magnetic reconnection in the Earth's magnetotail. Using a statistical sample of 140 reconnection outflows, we infer the acceleration potential associated with $E_\parallel$ from the shape of the electron velocity distribution functions. We show that heating by $E_\parallel$  in the reconnection outflow can reach up to ten times the inflow electron temperature. We demonstrate that the magnitude of the acceleration potential scales with the inflow Alfv\'en and electron thermal speeds to maintain quasi-neutrality in the reconnection region. Our results suggest that, as the inflow plasma parameter $\beta_{e\infty}$ increases, $E_\parallel$ becomes increasingly important to the ion-to-electron energy partition associated with magnetic reconnection.
\end{abstract}

\maketitle

\emph{Introduction} - Magnetic reconnection is a ubiquitous plasma process that powers some of the most extreme phenomena in the Universe, such as black hole flares, solar flares, and planetary magnetospheric storms~\cite{ripperda_black_2022,masuda_looptop_1994,angelopoulos_tail_2008}. Magnetic reconnection converts electromagnetic energy into ion and electron bulk and thermal energy through a topological reconfiguration of the magnetic field~\cite{yamada_magnetic_2010}. The reconnection electric field $E_{rec}$ and magnetic-field-aligned electric fields $E_\parallel$ have been identified as important agents in the electron acceleration and heating~\cite{le_twostage_2016,oka_particle_2023}. Recent \textit{in situ} observations~\cite{phan_electron_2013,oieroset_scaling_2023} and numerical simulations~\cite{shay_electron_2014,haggerty_competition_2015} show that the total electron heating by magnetic reconnection scales with the upstream electromagnetic energy per particle $m_iV_{Ai\infty}^2=B_\infty^2/\mu_0 n_\infty$, where $B_\infty$ is the asymptotic reconnecting magnetic field and $n_\infty$ is the asymptotic inflow particle number density. However, it remains unclear what controls the efficiency of the various heating mechanisms, particularly the acceleration by $E_\parallel$. Therefore, it is crucial to provide observational insights on electron heating by $E_\parallel$ to understand how energy is converted and distributed during collisionless magnetic reconnection.

The acceleration by $E_\parallel$ forms non-equilibrium electron velocity distribution functions (eVDFs), which consist of, e.g., a cold beam accelerated toward the X-line and a thermalized hot population~\cite{uzdensky_physical_2006,egedal_double_2015}. Such eVDFs trigger Buneman and electron two-stream instabilities, which excite electrostatic waves, reshaping the eVDF into a flat-top, i.e., constant phase-space density for electrons with parallel velocities $v_\parallel$ between thermal to supra-thermal speeds~\cite{asano_electron_2008,fujimoto_generation_2006,fujimoto_wave_2014,khotyaintsev_plasma_2011,khotyaintsev_electron_2020,norgren_electron_2020}. The energy of the accelerated cold beam corresponds to the net work done by the electric field along the electron path, i.e., the integrated electric field $E_\parallel$ along the field line, $e\Phi_\parallel(x) = -\int_\infty^x E_\parallel dl$ where $dl$ is a magnetic field line element~\cite{egedal_largescale_2012}. This acceleration (pseudo-)potential includes an electrostatic $\bm{E}=-\nabla \phi$ contribution and an inductive contribution $\bm{E}= -\partial \bm{A}/\partial t$~\cite{egedal_situ_2005}. \textit{In situ} observations and numerical simulations suggested that $E_\parallel$ are primarily electrostatic and maintain quasi-neutrality by supplying electrons and trapping them in the ion diffusion region (IDR)~\cite{egedal_situ_2005,egedal_double_2015,le_twostage_2016,wetherton_anisotropic_2021,lu_electrodynamic_2021}. However, it was also proposed that $E_\parallel$ is dominated by the electromagnetic reconnection electric field, which couples with the Hall magnetic field~\cite{bessho_electron_2015}. Therefore, the nature, role, and importance of the parallel electric fields in magnetic reconnection and the associated electron heating remain debated.

\emph{Data} - We use data from the Magnetospheric Multiscale (MMS) spacecraft \cite{burch_magnetospheric_2016} to investigate electron heating associated with magnetic-field-aligned electric fields $E_\parallel$ during magnetic reconnection in the Earth's magnetotail. In this region, the guide field ratio $B_g/B_\infty$ and the inflow $\beta_{e\infty}=2\mu_0 n_{\infty} k_B T_{e\infty} / B_{\infty}^2$ are typically small such as $B_g/B_\infty \lesssim 0.2$ and $\beta_{e\infty}\sim 0.001 - 0.1$~\cite{oieroset_scaling_2023}. We study eVDFs in 140 magnetic reconnection outflows in the plasma sheet of the Earth's magnetotail ($\beta_i \geq 0.5$, where $\beta_i = 2\mu_0 n_i k_B T_i / B^2$, $n_i$ is the ion number density, $T_i$ the ion temperature, and $B$ the magnetic field magnitude)~\cite{richard_are_2022}. We use magnetic field measurements from the FGM instrument \cite{russell_magnetospheric_2016}, the electric field from the EDP instrument~\cite{lindqvist_spinplane_2016,ergun_axial_2016}, ion moments from the HPCA instrument~\cite{young_hot_2016}, and electron VDFs measured by the FPI instrument \cite{pollock_fast_2016}. To improve the counting statistics and reduce the uncertainties due to low electron number densities ($n_e\leq 1~\mathrm{cm}^{-3}$)~\cite{gershman_calculation_2015}, we average $N = d_i / \langle V_i\rangle \tau_s \sim 100$ consecutive eVDFs, where $d_i=\sqrt{m_i/\mu_0 n_i e^2}$ is the ion inertial length, $\langle V_i\rangle$ the time-averaged ion bulk speed, and $\tau_s=30~\mathrm{ms}$ the FPI instrument's eVDF sampling time.\\

\begin{figure}[!t]
    \centering
    \includegraphics[width=\linewidth]{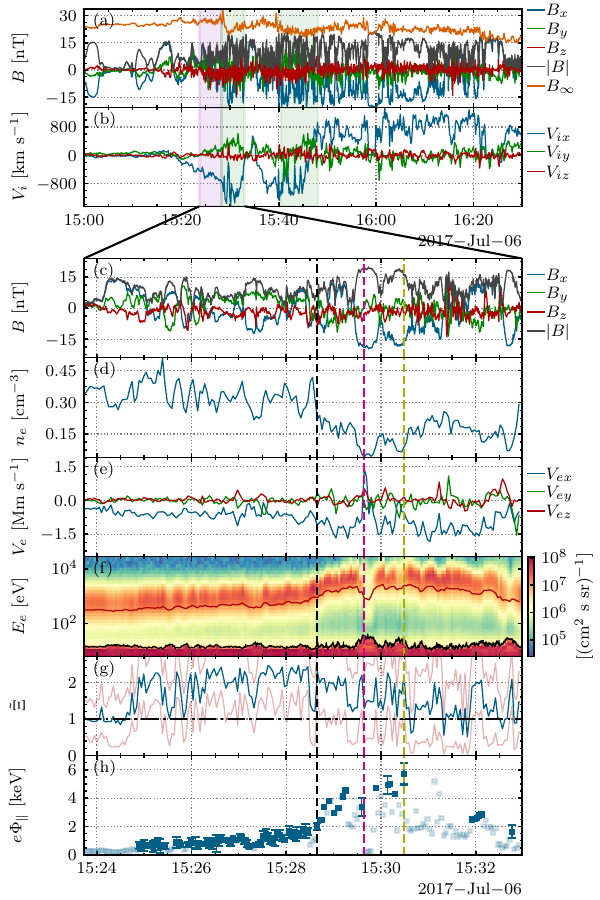}
    \caption{Example reconnection outflow. (a) and (c) Magnetic field in Geocentric Solar Magnetic (GSM) coordinates. (b) Ion bulk velocity. (d) Electron number density. (e) Electron bulk velocity. (f) Electron differential energy flux. (g) Normalized flatness factor $\tilde{\Xi}$. The blue line indicates the nominal value, and the red lines indicate $\tilde{\Xi}\pm \sigma(\tilde{\Xi})$,  where $\sigma(\tilde{\Xi})$ is the uncertainty propagated from $r$ and $q$. The black dashed-dotted line indicates the (bi-)Maxwellian level $\tilde{\Xi}=1$. (h) Acceleration potential. Shaded points in panel (h) indicate non-flat-top eVDFs, i.e., $\tilde{\Xi} - \sigma(\tilde{\Xi}) \leq 1$. The magenta and green shaded regions in panels (a)-(b) show the PSR and TLR intervals, respectively. The vertical dashed lines show the time of the eVDFs shown in Figs.~\ref{fig:vdfs}c-~\ref{fig:vdfs}e.}
    \label{fig:case}
\end{figure}

\begin{figure}[!t]
    \centering
    \includegraphics[width=\linewidth]{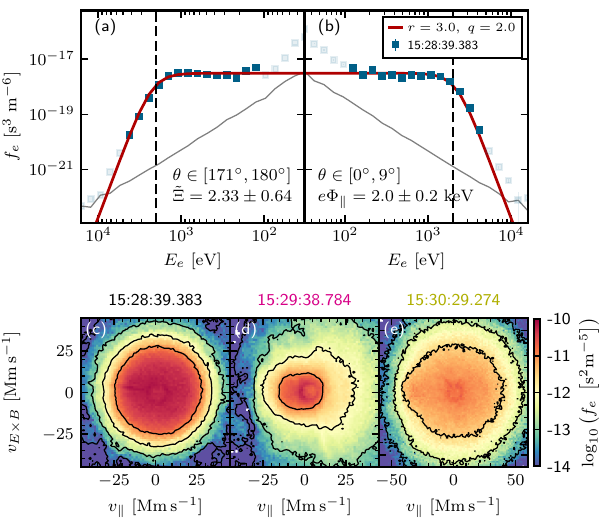}
    \caption{Examples of eVDFs in the plasma frame in the reconnection outflow. (a)-(b) Phase-space density of field-aligned electrons (blue squares) and $(r,q)$ fit (red). The error bars indicate the uncertainties due to counting statistics. Note that the error bars are smaller than the squares. The shaded squares correspond to energy bins not used in the fit, and the grey lines indicate the one-count level. The dashed lines indicate the estimated acceleration potential $e\Phi_\parallel$. (c)-(d) Reduced phase-space density in the $(v_\parallel, v_{E\times B})$ plane, taken from the times marked by the vertical lines in Fig.~\ref{fig:case}, corresponding to PSR, separatrix flow, and TLR respectively.}
    \label{fig:vdfs}
\end{figure}

\emph{Case study} - We present an example event in Fig.~\ref{fig:case}. Initially, the MMS spacecraft are in the quiet plasma sheet ($V_i \leq 100~\mathrm{km}~\mathrm{s}^{-1}$) [Fig.~\ref{fig:case}b] and observe large amplitude magnetic field oscillations [Fig.~\ref{fig:case}a] corresponding to flapping motions of the current sheet (CS)~\cite{richard_observations_2021}. During this time, the asymptotic magnetic field estimated from pressure balance $B_{\infty}= B\sqrt{1 + \beta_i}$~\cite{asano_evolution_2003} is stable. We see a fast tailward flow at 15:24:00 [Figs.~\ref{fig:case}b and~\ref{fig:case}e]. Then, from 15:28:30 to 15:33:00, the asymptotic magnetic field decreases, the speed of the outflow increases, the density drops, and the electron energy increases [Figs.~\ref{fig:case}a-~\ref{fig:case}f]. This suggests a transition from plasma sheet reconnection to tail lobe reconnection at 15:28:30~\cite{vaivads_suprathermal_2011}. At 15:29:40, which coincides with MMS approaching the inflow ($B\approx B_{\infty}$), we observe a separatrix electron flow toward the X-line~\cite{uzdensky_physical_2006,nagai_structure_2003,norgren_electron_2020}. The eVDF at the separatrix shows the accelerated beam and the hot thermalized electron population [Fig.~\ref{fig:vdfs}d]. Such eVDFs, which are commonly observed away from the reconnecting CS mid-plane~\cite{norgren_electron_2020,khotyaintsev_electron_2020}, can drive Buneman and electron two-stream instabilities producing flat-top eVDFs~\cite{fujimoto_generation_2006,fujimoto_wave_2014}. 

In the reconnection outflow, the (anti-)field-aligned electron phase-space density shows a clear flat-top shape with $\partial_{v_\parallel} f_e(v_\parallel, v_\bot=0) \approx 0$ below $2~\mathrm{keV}$ [Figs.~\ref{fig:vdfs}a and~\ref{fig:vdfs}b]. Note that the shaded points at low energies are due to contamination by spacecraft and instrument photo-electrons and should be ignored~\cite{gershman_spacecraft_2017}. We characterize the electron phase-space distribution using an $(r,q)$ model~\cite{qureshi_parallel_2004} 

\begin{equation}
    \label{eq:model}
    f_{(r, q)} = f_0\left [1 + \left (\frac{v_\parallel^2}{\xi v_{te,\parallel}^2} \right )^{r+1}\right ]^{-q},
\end{equation}

\noindent
with $f_0 = f(v_\parallel=0, v_\bot = 0)=n_e \eta/\pi^{3/2} v_{te\bot}^2v_{te\parallel}$ where $v_{te\parallel (\bot)}=\sqrt{2k_BT_{e\parallel(\bot)}/m_e}$ is the electron thermal velocity parallel (perpendicular) to the local magnetic field and

\begin{equation*}
    \eta = \frac{3\pi^{1/2} \Gamma(q)}{4\xi^{3/2}\Gamma(q-\mu)\Gamma(q+\mu)}, ~\xi=\frac{3\Gamma(\mu)\Gamma(q - \mu)}{2\Gamma(\nu)\Gamma(q - \nu)},
\end{equation*}

\noindent
where $\mu=3/2(r+1)$ and $\nu=5/2(r+1)$ with $r\in \mathbb{R}_+$ and $q\in ]1; +\infty[$. In Eq.~\ref{eq:model}, $q$ accounts for the high-energy tail of the eVDF and $r$ for the low-energy flat-top part, so that $f_{(r, q)}$ is a (bi-)Maxwellian for ($r=0$, $q\rightarrow +\infty$), a kappa distribution for ($r=0$, $q < +\infty$)~\cite{pierrard_kappa_2010}, and a flat-top for ($r>0$, $q < +\infty$). To estimate $r$, $q$ and their respective uncertainties, we fit the measured electron phase-space distribution to $f_{(r,q)}$ using a Levenberg-Marquardt algorithm~\cite{more_levenbergmarquardt_1978} to minimize the reduced weighted $\tilde{\chi}^2$~\cite{oka_electron_2022}. Since we focus on thermal electron heating, we use energy bins $E_e\leq 5 T_{e\parallel}$ to fit the eVDFs. Furthermore, as we are interested in heating by $E_\parallel$ we consider only pitch-angles in the spacecraft frame $\theta=\tan^{-1}\left (v_\bot / v_\parallel \right )$ within $9^\circ$ from $180^\circ$ [Fig.~\ref{fig:vdfs}a] and $0^\circ$ [Fig.~\ref{fig:vdfs}b]. We use a $9^\circ$ range of pitch angles as it is wide enough to ensure sufficient counting statistics, yet narrow enough to measure cold beams accurately [Fig.~\ref{fig:vdfs}c]. For the eVDF in Figs.~\ref{fig:vdfs}a and~\ref{fig:vdfs}b, we obtain $r=3.0$, $q=2.0$, which is significantly different from a (bi-)Maxwellian. To quantify how flat-top the eVDF is, we introduce the flatness factor $\Xi_{(r,q)}\equiv f_{(r,q)}(v_\parallel=v_{te\parallel}, v_\bot=0)/f_0$. Normalizing $\Xi_{(r,q)}$ to $\Xi_{bM}=1/e$, corresponding to $\Xi$ for a (bi-)Maxwellian, we obtain

\begin{equation}
    \tilde{\Xi} \equiv \frac{\Xi_{(r,q)}}{\Xi_{bM}}=e\left (1 + \frac{1}{\xi^{r+1}}\right )^{-q},
\end{equation}

\noindent
so that $0\leq \tilde{\Xi} \leq e$ and $\tilde{\Xi}=1$ for a (bi-)Maxwellian. We define a flat-top eVDF as $\tilde{\Xi} > 1 + e^{-1}$, where the e-folding threshold is chosen empirically based on a visual inspection. We also require that the uncertainty on $\tilde{\Xi}$ propagated from the uncertainties on $r$ and $q$ is such that $\tilde{\Xi} - \sigma(\tilde{\Xi}) > 1$ for flat-top eVDFs. For the example eVDF, $\tilde{\Xi}=2.32\pm 0.66> 1$, which is a clear flat-top [Figs.~\ref{fig:vdfs}a and~\ref{fig:vdfs}b]. Evaluating $\tilde{\Xi}$ for all eVDFs, we find that $41\%$ of eVDFs in the example reconnection outflow are flat-tops [Fig.~\ref{fig:case}g].

Since the flat-top eVDFs are shaped by the scattering of the accelerated beam by the electrostatic waves~\cite{fujimoto_wave_2014,khotyaintsev_electron_2020}, we use the high-energy cut-off of the plateau, i.e., the knee energy, as a proxy for the energy of the accelerated source beam, i.e., the acceleration potential $e\Phi_\parallel$~\cite{asano_electron_2008,egedal_largescale_2012}. We define the threshold phase-space density decay as $\varepsilon \equiv f_{(r,q)}(v_\parallel=v_\Phi, v_\bot=0)/f_0$ which yields the knee velocity

\begin{equation}
    v_\Phi = \left ( \varepsilon^{-1 / q} - 1\right )^{1/2(r+1)}\xi^{1/2}v_{te,\parallel}.
\end{equation}

\noindent Here, we choose $\varepsilon=1/e$, which qualitatively captures the high-energy cut-off of the plateau well. We note that $v_\Phi=v_{te,\parallel}$ for a (bi-)Maxwellian. For the eVDF in Figs.~\ref{fig:vdfs}a and~\ref{fig:vdfs}b, we obtain $e\Phi_\parallel = m_e v_\Phi^2 / 2e = 2.0\pm 0.3~\mathrm{keV}$ (dashed lines).

Thereafter, for all flat-top eVDFs, we estimate the acceleration potential $e\Phi_\parallel$ [Fig.~\ref{fig:case}h]. The acceleration potential increases substantially from the plasma sheet reconnection (PSR) to tail lobe reconnection (TLR) regimes $\langle e\Phi_{\parallel}^{(TLR)}\rangle / \langle e\Phi_{\parallel}^{(PSR)} \rangle \approx 3.2$. To compare the change in $e\Phi_\parallel$ with the change in the plasma conditions, we estimate the inflow conditions for PSR and TLR at 15:01:49 and 15:59:24, respectively. We choose the inflows at $t_\infty = \operatorname{argmax}(B(t)/ B_\infty)$, so that $n_\infty \approx n(t_\infty)$ and $B_\infty \approx B(t_\infty)$, when the spacecraft are in the quiet plasma sheet and the lobe. We obtain $V_{Ae\infty}^{(PSR)}=29\times 10^6~\mathrm{m}~\mathrm{s}^{-1}$ and $V_{Ae\infty}^{(TLR)}=91\times 10^6~\mathrm{m}~\mathrm{s}^{-1}$, where $V_{Ae\infty}=B_\infty / \sqrt{\mu_0 n_{\infty}m_e}$ is the inflow electron Alfv\'en speed. These values yield, $V_{Ae\infty}^{(TLR)} / V_{Ae\infty}^{(PSR)}\approx 3.1\sim \langle e\Phi_{\parallel}^{(TLR)}\rangle / \langle e\Phi_{\parallel}^{(PSR)} \rangle$, which indicates that the acceleration potential increases with the inflow Alfv\'en speed $V_{Ae\infty}$.

To clarify the nature of the parallel electric fields, we use the electron momentum balance and investigate the changes of $e\Phi_\parallel$ across the reconnection outflow. We see that as the spacecraft moves away from the CS center at 15:25:44 to the CS edge at 15:27:15 corresponding to minimum and maximum of $B/B_\infty$ in the PSR outflow respectively, the acceleration potential drops down by $e\Delta \Phi_\parallel^{obs}=e \Phi_\parallel^{center} - e \Phi_\parallel^{edge}\approx 220~\mathrm{eV}$ [Fig.~\ref{fig:case}h]. From electron momentum balance along the field-line, neglecting the electron inertia and assuming gyrotropic electrons $\bm{P}_e = p_{e\bot} \bm{1} + (p_{e\parallel} - p_{e\bot})\bm{b}\bm{b}$ with $\bm{b}=\bm{B}/|\bm{B}|$, we obtain 

\begin{equation}
    -eE_\parallel = \nabla_\parallel T_{e\parallel} + T_{e\parallel}\nabla_\parallel \ln n_e + \left ( T_{e\bot} - T_{e\parallel} \right )\nabla_\parallel \ln B,
    \label{eq:momentum-balance}
\end{equation}

\noindent
where $\nabla_\parallel=\bm{b}\cdot \nabla$ is the gradient along the field line. Integrating Eq.~\ref{eq:momentum-balance} between two points $a$ and $b$, we obtain $e\Delta \Phi_\parallel = e\Phi_\parallel^{(a)} - e\Phi_\parallel^{(b)} = -e\int_b^a E_\parallel dl \simeq e\Delta \Phi_{\parallel T} + e\Delta \Phi_{\parallel n} + e\Delta \Phi_{\parallel B}$ with the electron temperature gradient contribution $e\Delta \Phi_{\parallel T} = T_{e\parallel}^{(a)} - T_{e\parallel}^{(b)}$, the density gradient contribution $e\Delta \Phi_{\parallel n} = T_{e\parallel}^{(a)} \ln \left ( n_e^{(a)}/n_e^{(b)}\right )$, and the magnetic field gradient $e\Delta \Phi_{\parallel B} = ( T_{e\bot}^{(a)} - T_{e\parallel}^{(a)}) \ln \left ( B^{(a)}/B^{(b)} \right )$~\cite{haggerty_competition_2015}. We take $a$ and $b$ as the CS center and edge, respectively, and assume the two measurement points are along the same field line. While this assumption may not be strongly verified, it provides a useful order-of-magnitude estimate. We obtain $e\Delta \Phi_{\parallel T}\approx 10~\mathrm{eV}$, $e\Delta \Phi_{\parallel B}\approx 50~\mathrm{eV}$, and $e\Delta \Phi_{\parallel n}\approx 160~\mathrm{eV}$ so that $e\Delta \Phi_\parallel\approx 220~\mathrm{eV}\sim e\Delta \Phi_\parallel^{obs}$. This suggests that the field-aligned ambipolar electric field is primarily due to electron density gradients~\cite{haggerty_competition_2015}.

\begin{figure}[!t]
    \centering
    \includegraphics[width=\linewidth]{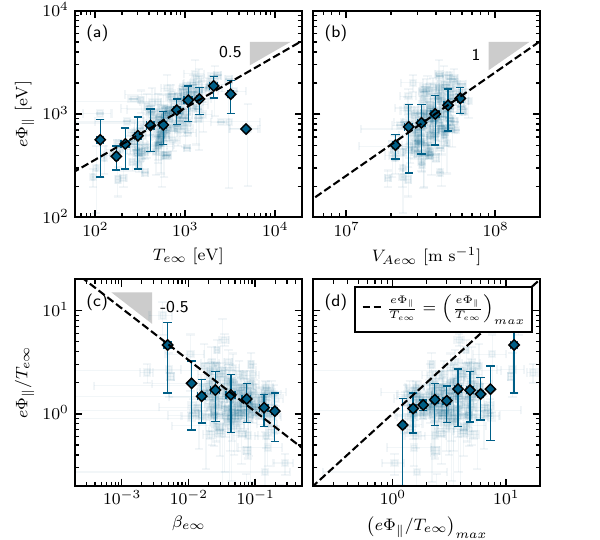}
    \caption{Statistics of the acceleration potential in the reconnection outflows. (a), (b) Acceleration potential as a function of the inflow electron Alfv\'en speed $V_{Ae\infty}$ and inflow electron temperature $T_{e\infty}$, respectively. (c), (d) Normalized acceleration potential versus the inflow $\beta_{e\infty}$, and the predicted $\left (e\Phi_\parallel/T_{e\infty}\right )_{max}$ computed using Eq.~\ref{eq:phi-max}. Transparent squares correspond to individual cases with the corresponding standard deviation, and solid diamonds correspond to the binned average and standard deviation. The black dashed lines in panels (a)-(c) indicate scaling as theoretically predicted by Ref.~\cite{le_magnitude_2010}.}
    \label{fig:blobs}
\end{figure}

\emph{Statistical results} - Figure~\ref{fig:blobs} shows the acceleration potential obtained for all 140 reconnection outflows. We select the inflow, as described in the case study, for each event to obtain the asymptotic parameters. Events that showcase both PSR and TLR, similar to the example in Fig.~\ref{fig:case}, are divided into two cases. First, we focus on the effect of the inflow temperature $T_{e\infty}$ on the acceleration potential. Recent studies suggested that a larger inflow temperature $T_{e\infty}$ increases the number of demagnetized electrons, which helps to maintain quasi-neutrality, thus reducing the parallel electric field~\cite{nan_formation_2022,lu_electrodynamic_2021}. However, we find that the acceleration potential increases as $e\Phi_\parallel\propto T_{e\infty}^{1/2}$ [Fig.~\ref{fig:blobs}a].

We find that the acceleration potential increases as $e\Phi_\parallel\propto V_{Ae\infty}$ [Fig.~\ref{fig:blobs}b] in agreement with the example case. By solving the electron current sheet firehose stability condition, $\Lambda = \mu_0(p_\parallel - p_\bot)/B^2 = 1$
for adiabatically trapped electrons in the reconnection inflow, Ref.~\cite{le_magnitude_2010} derived that the acceleration potential follows the scaling $e\Phi_\parallel/T_{e\infty} \propto \beta_{e\infty}^{-1/2}$ or equivalently, $e\Phi_\parallel \propto T_{e\infty}^{1/2}V_{Ae\infty}$. Assuming that $T_{e\infty}$ is independant of $V_{Ae\infty}$, we therefore expect $e\Phi_\parallel \propto T_{e\infty}^{1/2}$ and $e\Phi_\parallel \propto V_{Ae\infty}$, consistent with our observations [Figs.\ref{fig:blobs}a and~\ref{fig:blobs}b]. Moreover, the combined scaling $e\Phi_\parallel/T_{e\infty} \propto \beta_{e\infty}^{-1/2}$
predicted by Ref.\cite{le_magnitude_2010} agrees with our observations within the measurement and statistical uncertainties [Fig.\ref{fig:blobs}c].

To elucidate the role of the parallel electric fields, we compare $e\Phi_\parallel/T_{e\infty}$ against the theoretical prediction derived in Ref.~\cite{le_magnitude_2010}. The fluid firehose stability condition, i.e., global electron momentum balance of the electron current sheet~\cite{egedal_force_2023}, requires that the pressure anisotropy set by trapping electrons in the inflow is small $p_\parallel - p_\bot \leq B^2 / \mu_0$. The maximal acceleration potential is expected just upstream of the electron diffusion region, corresponding to the maximum pressure anisotropy, i.e., $\Lambda = 1$. Its scaling was derived in Ref.~\cite{le_magnitude_2010} and is given by

\begin{equation}
    \label{eq:phi-max}
    \left (\frac{e\Phi_\parallel}{T_{e\infty}} \right )_{max} \simeq \frac{1}{2} \left [ \left (\frac{4}{ \beta_{e\infty}} \right )^{1/4} - \frac{1}{2}\right ]^2.
\end{equation}
\noindent
We find that $e\Phi_\parallel/T_{e\infty} \leq \left (e\Phi_\parallel/T_{e\infty} \right )_{max}$ [Fig.~\ref{fig:blobs}d], suggesting that the fluid firehose threshold $\Lambda=1$ is not reached. This implies that other processes such as kinetic firehose~\cite{le_threedimensional_2019,cozzani_direct_2023} and current sheet scattering~\cite{wang_electron_2016} limit the pressure anisotropy.

\emph{Discussion} - Our statistical results show that the acceleration potential increases with the inflow Alfv\'en speed $V_{Ae\infty}$ [Fig.~\ref{fig:blobs}b]. We interpret this as a result of an increased parallel electron flux $n_e v_{\Phi}$ in the high Alfv\'en speed regime. The parallel electron flux, which supplies electrons in the IDR to maintain the quasi-neutrality, is typically explained as follows~\cite{uzdensky_physical_2006}: When a flux tube in the inflow convects into the IDR, its volume increases by the conservation of magnetic flux. Since ions are demagnetized, their density is approximately constant $n_i \simeq n_\infty$~\cite{egedal_double_2015} whereas, in the absence of electron flux, the electron density must decrease as $n_e/n_\infty=B/B_\infty$ which can be written as $n_e/n_\infty=V_{Ae}^2/V_{Ae\infty}^2$. To prevent such macroscopic charge density, parallel electric fields outside the IDR pull electrons in and trap them inside the IDR~\cite{egedal_situ_2005,uzdensky_physical_2006,egedal_double_2015}. Therefore, a larger inflow Alfv\'en speed $V_{Ae\infty}$, which results in a larger flux tube expansion rate, requires a larger electron flux to maintain quasi-neutrality and, thus, a larger acceleration potential.

 We also find that the acceleration potential increases with the inflow temperature $T_{e\infty}$. Our interpretation is that a larger $T_{e\infty}$ allows more electrons to overcome the potential barrier and escape. Therefore, the acceleration potential $e\Phi_\parallel$ must increase with $T_{e\infty}$ to keep electrons trapped in the IDR to maintain quasi-neutrality. We find that the acceleration potential increases as $e\Phi_\parallel\propto T_{e\infty}^{1/2}$ [Fig.~\ref{fig:blobs}a] which supports our interpretation.

We examined symmetric magnetic reconnection in the Earth's magnetotail, where the guide field is weak. Numerical simulations and \textit{in situ} observations confirm similar acceleration potential scaling laws for asymmetric reconnection with a strong guide field, like at the dayside magnetopause~\cite{egedal_electron_2011,graham_electron_2014}. However, in asymmetric reconnection, $e\Phi_\parallel$ varies between inflows due to differences in density and magnetic field strength, requiring different trapping potential to maintain quasi-neutrality. Since cold magnetosheath electrons dominate the outflow of dayside reconnection, applying our method to dayside reconnection would therefore mainly probe $e\Phi_\parallel$ of the cold dense inflow.

Recent \textit{in situ} observations and numerical simulations showed that the net electron heating between the reconnection inflow and the outflow empirically scales as $\Delta T_e=\alpha_e m_iV_{Ai\infty}^2$ with $\alpha_e\sim 0.02$~\cite{phan_electron_2013,shay_electron_2014,haggerty_competition_2015,le_twostage_2016,oieroset_scaling_2023}. Similarly, the ion heating is given by $\Delta T_i=\alpha_{i1}m_iV_{Ai\infty}^2 - 2\alpha_{i2}e\Phi_\parallel$, where $\alpha_{i1}=0.13$ and $\alpha_{i2}=3/4$ were obtained empirically~\cite{oieroset_scaling_2024}. The first term corresponds to the pick-up acceleration of ions~\cite{drake_ion_2009} while the second term is due to the deceleration of ions by $E_\parallel$~\cite{haggerty_competition_2015}. Combining with the scaling $e\Phi_\parallel / T_{e\infty} = \alpha_\Phi \beta_{e\infty}^{-1/2}$, where $\alpha_\Phi\approx 0.31$ from our estimates of $e\Phi_\parallel$ [Fig.~\ref{fig:blobs}c], we obtain

\begin{equation}
    \frac{\Delta T_i}{\Delta T_e} = \frac{\alpha_{i1}}{\alpha_e} \left ( 1 - \frac{\alpha_{i2}\alpha_\Phi}{\alpha_{i1}} \sqrt{\beta_{e\infty}} \right ).
    \label{eq:ion-to-electron}
\end{equation}

\noindent
In the limit of $\beta_{e\infty} \ll 1$, $\Delta T_i/\Delta T_e \simeq \alpha_{i1}/\alpha_e \approx 6.5$, and for $\beta_{e\infty}\simeq (\alpha_{i1} / \alpha_{i2}\alpha_\Phi)^2\approx 0.3$,  $\Delta T_i \ll \Delta T_e$. For magnetotail conditions $\beta_{e\infty}=0.001-0.1$, we obtain $\Delta T_i/\Delta T_e = 2.9-6.1$, consistent with statistically observed values~\cite{oieroset_scaling_2024}. Thus, Eq.~\ref{eq:ion-to-electron} suggests that, as the inflow plasma parameter $\beta_{e\infty}$ increases, $E_\parallel$ becomes increasingly important to the ion-to-electron energy partition.

\emph{Conclusions} - We presented a statistical study of electron heating by magnetic-field-aligned electric fields in magnetotail reconnection outflows. We determine the acceleration potential, i.e., the net work done by the electric field on the electrons, based on the analysis of the eVDFs in the reconnection outflow. We find that the acceleration potential can reach up to ten times the inflow temperature $T_{e\infty}$ and scales as $e\Phi_\parallel \propto T_{e\infty}^{1/2}V_{Ae\infty}$ necessary to maintain quasi-neutrality. Our results emphasize the importance of parallel electric fields in electron heating in magnetic reconnection.

Combining the observed scaling with empirical ion and electron heating relations, we show that the ion-to-electron heating ratio decreases as $\Delta T_i/ \Delta T_e \propto 1-\beta_{e\infty}^{1/2}$. This result can help to interpret remote observations where detailed plasma properties cannot be measured. For example, estimating the electron density and the magnetic field from radio images of accretion flows requires an assumption on $T_i/T_e$~\cite{chael_role_2018,theeventhorizontelescopecollaboration_first_2019}.

\begin{acknowledgments}
MMS data are available at the MMS Science Data Center~\footnote{\url{https://lasp.colorado.edu/mms/sdc/public}.}. Data analysis was performed using the \verb+pyrfu+ analysis package~\footnote{\url{https://pypi.org/project/pyrfu/}.}. 
The dataset used to generate Fig.~\ref{fig:blobs} is publicly available from the
Zenodo repository \footnote{See \url{https://doi.org/10.5281/zenodo.14984107}}. We thank the MMS team and instrument PIs for data access and support. The work was supported by the Knut and Alice Wallenberg Foundation (Dnr. 2022.0087).
\end{acknowledgments}

\bibliographystyle{apsrev4-2}
\bibliography{main}

\end{document}